# Early warning in egg production curves from commercial hens: A SVM approach.

Iván Ramírez Morales*[a,b], Daniel Rivero Cebrián[b], Enrique Fernández Blanco[b], Alejandro Pazos Sierra[b]

[a] **Universidad Técnica de Machala,** Faculty of Agricultural & Livestock Sciences. Address: 5.5 km Pan-American Av, Machala, El Oro, Ecuador.

[b] **Universidade A Coruña,** Department of Computer Science. Address: 15071 A Coruña A Coruña (03082), A Coruña, España.

**\* Corresponding Author:** iramirez@utmachala.edu.ec

**Early warning in egg production curves from commercial hens: A SVM approach.**

**Abstract**

Artificial Intelligence allows the improvement of our daily life, for instance, speech and handwritten text recognition, real time translation and weather forecasting are common used applications. In the livestock sector, machine learning algorithms have the potential for early detection and warning of problems, which represents a significant milestone in the poultry industry. Production problems generate economic loss that could be avoided by acting in a timely manner.

In the current study, training and testing of support vector machines are addressed, for an early detection of problems in the production curve of commercial eggs, using farm´s egg production data of 478,919 laying hens grouped in 24 flocks.

Experiments using support vector machines with a 5 k-fold cross-validation were performed at different previous time intervals, to alert with up to 5 days of forecasting interval, whether a flock will experience a problem in production curve. Performance metrics such as accuracy, specificity, sensitivity, and positive predictive value were evaluated, reaching 0-day values of 0.9874, 0.9876, 0.9783 and 0.6518 respectively on unseen data (test-set).

The optimal forecasting interval was from zero to three days, performance metrics decreases as the forecasting interval is increased. It should be emphasized that this technique was able to issue an alert a day in advance, achieving an accuracy of 0.9854, a specificity of 0.9865, a sensitivity of 0.9333 and a positive predictive value of 0.6135. This novel application embedded in a computer system of poultry management is able to provide significant improvements in early detection and warning of problems related to the production curve.

1.  **Introduction**

Poultry farmers have used data to monitor health and production of their flocks for over 40 years. Data such as consumption of food, water, growth and mortality have been collected in order to monitor and improve yields, and these data and statistics are plotted on a graph and used as early indicators of the health and welfare of poultry (Hepworth et al., 2012).

Egg producers usually know and record the number of eggs produced, frequently a production curve is plotted and monitored in order to detect problems in the production curve indicating a possible disease, or any other issues (Long and Wilcox, 2011).

The curve of egg production can be affected by various factors such as food intake (quality and quantity), water consumption, intensity and duration of the light received, vermin infestation, diseases and other handling or environmental causes (Jacob et al., 2014).

When it comes to a disease, having early detection tools is of vital importance. That is, before it is spreading to other animals and/or becoming entrenched in the environment. The early detection of a problem means acting in a timely manner; reducing the cost and increasing the effectiveness of the treatment or control of a disease are directly related to the time it takes to detect it (Schaefer et al., 2004; Cameron, 2012).

The machine learning algorithms are present in various activities of our daily life, and they allow discovering rules and patterns in data sets. For example, in epidemiology, the supervised machine learning has the potential to classify, diagnose and identify risks. Support vector machines, are one of this algorithms, the main feature is that they can learn how to classify data from examples (McQueen et al., 1995; Hepworth et al., 2012).

References to studies that used machine learning techniques in livestock have been found, for example, various algorithms were employed to predict the rate of pregnancy, or weight in cattle, from routine production data (Hempstalk et al., 2015; Alonso et al., 2015).

Support vector regression and neural networks to predict the body and carcass characteristics of broilers (Faridi et al., 2012). Support vector machines to predict hock burn in chickens (Hepworth et al., 2012). Artificial intelligence and images to detect the avian smallpox (Hemalatha et al., 2014).

Lokhorst and Lamaker, (1996) reported an expert system for monitoring the daily production process in aviary systems for laying hens, however, no information has been found regarding the early detection of problems using farm´s data which are normally recorded in poultry production.

To the best of the authors' knowledge, there are no prior studies on using machine learning algorithms for early detection of problems in the egg production curve from commercial hens. Although, since the early 1980s there are similar works in the mathematical study of the production curve of laying hens. Nonlinear models have been widely used to adjust the curves of egg production in laying hens (Adams and Bell, 1980; Grossman and Koops, 2001; Savegnago et al., 2012).

Moreover, a vast amount of literature has been compiled, for over 30 years, on the use of control charts to monitor animal farming, but its practical use does not seem to be widespread (De Vries and Reneau, 2010).

Studies such as those carried out by Grossman et al., (2000) and Narinc et al., (2014) have been found, who developed mathematical models to describe the production curve and the persistence of the curve in laying hens. Other works, such as those published by Long and Wilcox, (2011), studied the production curve of laying hens to determine whether the economic use of flocks of laying hens was optimal.

Some learning techniques have been used to model the production curve, especially artificial neural networks, showing that they are able to successfully replace traditional mathematical and statistical models when predicting egg production in laying hens. These models, which are easier to use, require fewer variables and can be more efficiently compared with their mathematical counterparts (Ahmadi and Golian, 2008; Ahmad, 2011; Felipe et al., 2015).

There is general agreement on the need to monitor the production yield of farm animals, that is why mathematical methods (Dohoo, 1993), recursive algorithms (Roush et al., 1992), data display systems and statistical techniques (Woodall and Tech, 2006) have been used. Significant differences indicating an alteration in the productive indicators of farm animals are sought (De Vries and Reneau, 2010).

The real-time monitoring is a major challenge because data collection includes natural variability; Woudenberg et al., (2014) developed a method for early detection of problems based on the calculation of waste, which allows identifying potential problems in egg production from 10 production flocks.

The concept of control charts as part of the statistical process control is commonly used to monitor industrial processes; several authors demonstrated their use in the context of animal husbandry, although the statistical properties of data regarding animals often do not meet the basic principles of these control charts (Mertens et al., 2011).

In the above-mentioned cases, computer-aided detection methods were presented, but no publications were found on the use of machine learning algorithms aimed at developing models that allow partial automation of this task.

This study is aimed at developing and testing an early warning model based on support vector machines algorithms, in order to detect problems in egg production curve from commercial hens.

## 2. Materials and methods

*2.1 Data description*

A farm database of egg production of laying hens of the ISA Brown, Lohmann Brown and H&N layer lines were used, collected over a period of seven years (January 2008 to December 2014) from a poultry company. Data correspond to 24 flocks, of approximately 20,000 birds at the beginning of the production cycle, using the "all-in all-out" replacement system, i.e. each flock contains only birds of the same age at the beginning, during the entire production period and when the production cycle is completed.

Data are recorded once a day, at the end of the day, but not always at same time, it is done when counting and sorting of eggs and dead birds have been carried out, and it also depends on the weekday. The production period used for the experiments encompassed 60 weeks (from age 19 to 79 weeks), for each day in which there was a production problem was labeled as positive by an experts' panel formed by the farm's production manager (veterinarian), the owner who has been poultry farmer for 30 years, and a local poultry veterinarian.

The average number of days labeled as positive for each flock is 8 days, however, it is observed that there are flocks, which present no problem, and there are others, which present up to 33 days labeled as positive. In total, the 24 flocks, throughout the 7 years of study, pre-sented 188 positive labels, representing only 1.85% of the 10,142 records. That is, the classifier has a lot of negative patterns (days when there are no problems) and few positive patterns, this fact unbalances the expected outputs and adds difficulty to the task of classification and forecasting.

*Table 1 Main production indicators of the flocks under study.*

| Flock | Production time (d) | Housed birds | Dead birds | Total amount of eggs | Average eggs per day | Eggs/housed bird/day | Peak of the production % per bird/day | Positive labels |
|---|---|---|---|---|---|---|---|---|
| 1 | 473 | 20,300 | 2,929 | 7,211,252 | 15,246 | 0.7510 | 98.14% | 30 |
| 2 | 429 | 20,361 | 3,022 | 6,951,132 | 16,203 | 0.7958 | 97.04% | 0 |
| 3 | 516 | 20,137 | 3,630 | 8,160,013 | 15,814 | 0.7853 | 96.87% | 1 |
| 4 | 148 | 18,874 | 1,430 | 1,767,577 | 11,943 | 0.6328 | N/A | 0 |
| 5 | 480 | 19,770 | 2,421 | 7,185,831 | 14,970 | 0.7572 | 97.25% | 0 |
| 6 | 461 | 20,408 | 1,573 | 7,145,492 | 15,500 | 0.7595 | 97.11% | 33 |
| 7 | 518 | 20,187 | 2,718 | 7,974,633 | 15,395 | 0.7626 | 95.97% | 14 |
| 8 | 501 | 20,130 | 1,984 | 8,093,083 | 16,154 | 0.8025 | 97.03% | 0 |
| 9 | 104 | 19,740 | 436 | 1,594,527 | 15,332 | 0.7767 | 97.03% | 0 |
| 10 | 389 | 19,668 | 2,153 | 6,078,320 | 15,626 | 0.7945 | 95.86% | 0 |
| 11 | 543 | 19,920 | 2,409 | 7,900,793 | 14,550 | 0.7304 | 97.32% | 0 |
| 12 | 491 | 19,934 | 1,969 | 7,230,558 | 14,726 | 0.7387 | 98.70% | 17 |
| 13 | 431 | 19,492 | 1,382 | 6,787,937 | 15,749 | 0.8080 | 96.30% | 13 |
| 14 | 419 | 19,920 | 1,600 | 7,147,832 | 17,059 | 0.8564 | 97.17% | 0 |
| 15 | 468 | 20,120 | 1,549 | 7,172,119 | 15,325 | 0.7617 | 98.91% | 0 |
| 16 | 517 | 20,234 | 2,865 | 7,692,698 | 14,879 | 0.7354 | 97.42% | 12 |
| 17 | 498 | 19,971 | 2,051 | 7,744,766 | 15,552 | 0.7787 | 96.83% | 24 |
| 18 | 391 | 20,104 | 1,238 | 6,463,994 | 16,532 | 0.8223 | 97.72% | 13 |
| 19 | 307 | 20,094 | 693 | 5,301,566 | 17,269 | 0.8594 | 98.52% | 0 |
| 20 | 450 | 19,895 | 1,984 | 6,905,452 | 15,345 | 0.7713 | 98.16% | 13 |
| 21 | 480 | 19,910 | 2,702 | 7,590,784 | 15,814 | 0.7943 | 96.94% | 0 |
| 22 | 529 | 19,950 | 2,973 | 8,429,271 | 15,934 | 0.7987 | 98.20% | 10 |
| 23 | 374 | 19,907 | 2,050 | 6,023,519 | 16,106 | 0.8090 | 98.30% | 8 |
| 24 | 202 | 19,893 | 814 | 3,407,626 | 16,869 | 0.8480 | 97.63% | 0 |

Table 1 describes each flock with its corresponding general indicators: production time; birds housed at the beginning of the production cycle; dead birds during the production time; total eggs produced by the flock during the production time; average number of eggs produced per day; daily eggs per hen housed; production maximum % (peak) reached and the number of positive labels of each flock.

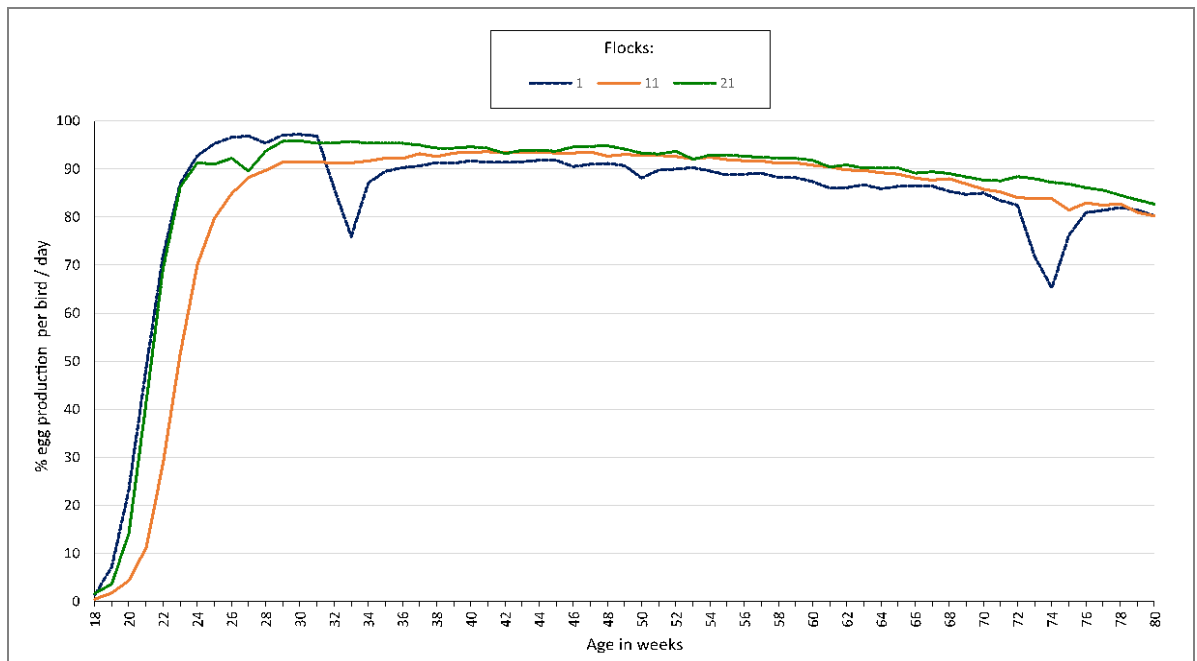

*Figure 1 Weekly average production per bird in three representative flocks.*

Figure 1 shows three flocks which are representative for the database: the solid line represents flock 11, which has a characteristic curve, without any problems throughout the production time; the dotted line represents flock 21, which has small drops and delays in the production curve, but they are not significant; the dashed line represents flock 1, which has two significant production drops, the first one begins near 31 weeks and the second at 72 weeks old.

From the numerous meetings with the poultry farmers, it is found that on farms where the collection of eggs is done at a specific time, there is greater data consistency than on those where it is carried out at different times each day; on the farm where no standard time routine was established for the collection of eggs, in either house, the number of eggs produced per day varies. This variability can be observed in Figure 2, representing the daily egg production per bird. This fact represents an additional challenge for the early warning model, because it should be able to distinguish between a real problem and these drops due to weekly cyclical variations related to routine and time of collection

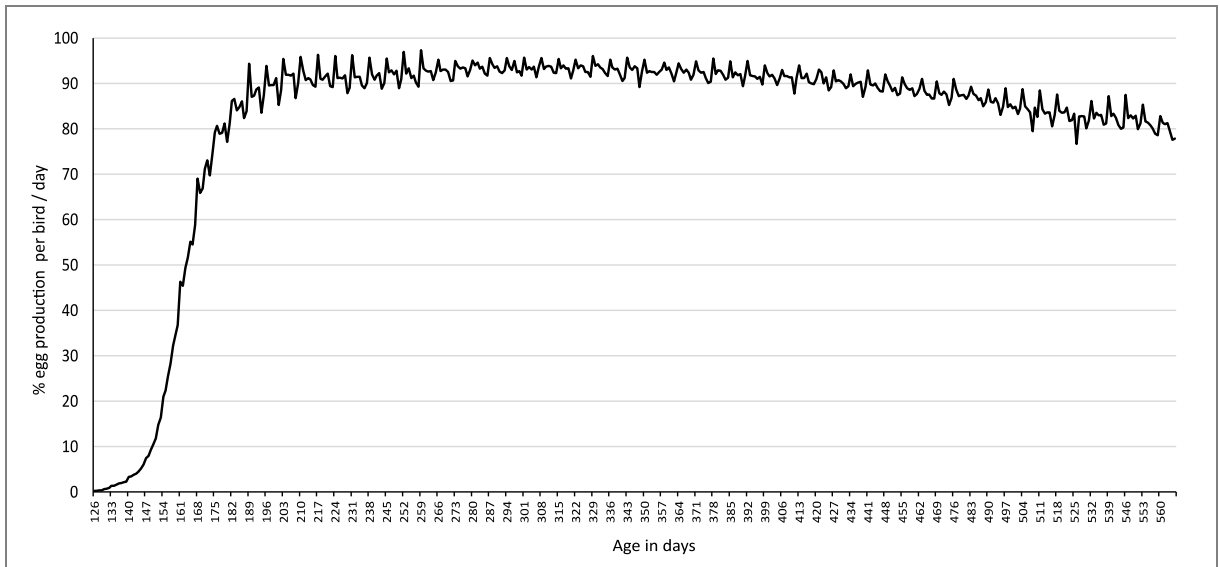

*Figure 2 Daily records of egg production per bird in flock 11.*

Figure 3 shows an example of problem zone tagged using poultry experts' judgment, each production day of each flock was labeled with values of 0 in the absence of a problem and 1 otherwise.

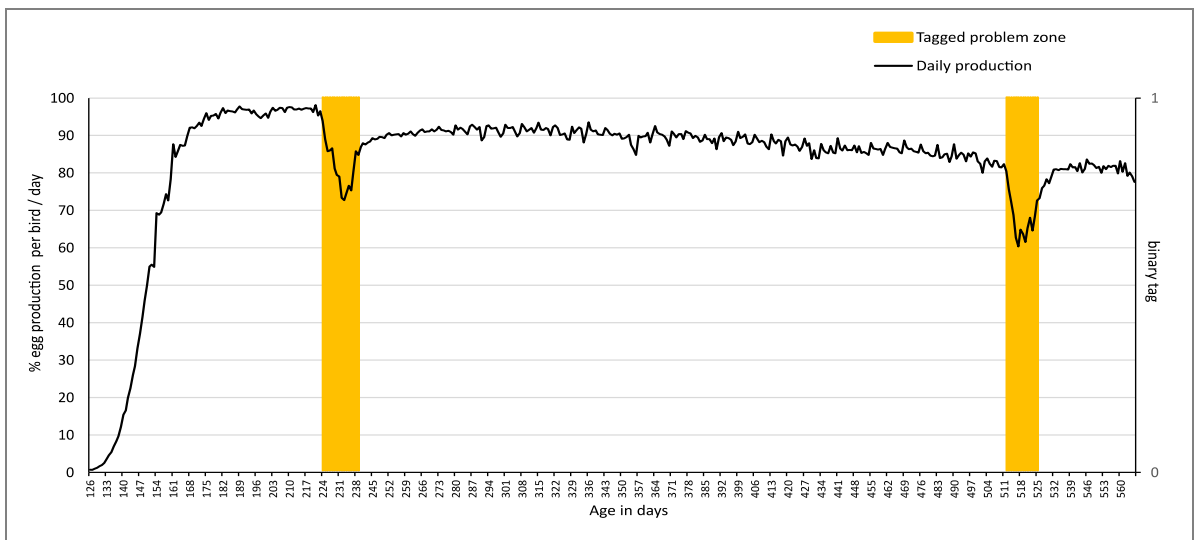

*Figure 3. Labeling example of problems in flock 1.*

## 2.2 Support Vector Machines algorithms

Basically there are two types of machine learning algorithms: supervised and unsupervised; the former is used when there is knowledge about the desired outputs, and it is trained to obtain them, whereas the latter generates a grouping (clusters) without information on the expected outputs (Mucherino et al., 2009).

Once an algorithm has been trained, it is able to transfer the learned dependence between the input patterns (features) and expected outputs (targets) into new data. The quality of a classifier can be measured by the proportion of correctly classified patterns in the test set, i.e. in new data that were not used during training, this set allows for assessing the error in the generalization of the final model chosen (Hastie et al., 2009).

Among the most commonly used techniques for data mining, are the support vector machines (SVM), which are supervised machine learning algorithms used to classify data sets into two different classes, separated by a hyperplane defined in an appropriate space (Mucherino et al., 2009).

They can be used in classification and regression problems, as their functioning starts from a set of training samples whose classes are labeled, and they also train an SVM to build a model that predicts the class of a new sample, different from the original one (Palma and Marín, 2013; Benítez et al., 2013).

The basics of SVM were developed by Vapnik and Chervonenkins in 1963, in a study on the theories of statistical learning that was aimed at narrowing down the generalization error according to the complexity of the search space. In 1992 Vapnik, Boser and Guyon proposed a method to create non-linear classifiers (Boser et al., 1992), and the current standard of SVM was proposed by Cortes and Vapnik, (1995). The purpose of SVM is for obtaining models which structurally have little risk of error regarding future data. Although originally they were designed to solve binary (two classes) classification problems, their application has been extended to regression, multiclassification, clustering and other tasks (Palma and Marín, 2013).

This technique is intended to find an optimal hyperplane able to distribute data into the classes to which they belong. Intuitively, it seems obvious to conclude that when facing a problem of linear classification there is a high probability of obtaining several solutions which successfully classify data (Fernandez-Lozano et al., 2013).

The optimal hyperplane used to separate the two classes can be defined from a small amount of data from the training set called support vectors, which determine the margin (Cortes and Vapnik, 1995; Mucherino et al., 2009). Figure 4 shows the above-mentioned concepts.

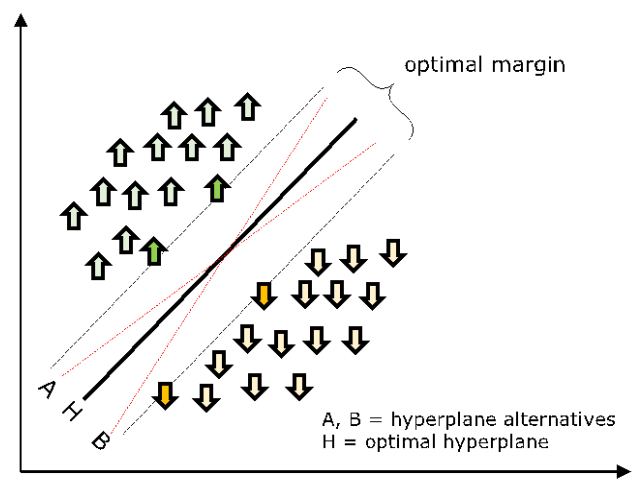

*Figure 4 A problem separable in a two-dimensional space. Support vectors define the margin of greatest separation between classes.*

The choice of the best hyperplane was solved in 1965 (Vapnik and Kotz, 1982) with the approach that the optimal hyperplane is defined as the linear decision function with the maximum margin between the vectors of the two classes.

However, in most problems, the data are not linearly separable and it is required to use strategies such as the identification of other separation dimensions. The kernel functions are used to transform the original multidimensional space into another, where classes are linear-ly separable. In practice, support vector machines are trained using different kernels to select the one with the best performance for the problem raised (Mucherino et al., 2009).

Some preliminary tests were performed by trial and error on the test set (Mollazade et al., 2012), using most common available kernels, however only four had an acceptable accuracy to the opinion of the authors. The research focused on these kernels: polynomial, radial basis function – RBF (Gaussian), quadratic and linear, in order to perform an exhaustive evaluation.

The polynomial and RBF kernels are among the most commonly used ones; the latter has a sigma (σ) parameter which can be tuned to adjusts the size of the kernel (Bennett and Campbell, 2000). Preliminary tests were performed to select the sigma tuning best range, which was between one and six; this range was used for exhaustive evaluation.

SVM has a compensation parameter C, which can be modified and affects the classification quality, since it determines how severely any misclassification should be penalized; generally, very high C values may lead to overfitting problems, reducing the SVM ability to generalize (Mucherino et al., 2009). In order to evaluate this parameter without overfitting the classifier, values below 0.25 were selected.

*2.3 Data processing*

Starting from the production data, two sets of patterns were created: the inputs, which had SVM and the desired outputs for them. The input patterns are made up by taking data from a sliding window (Lindsay and Cox, 2005), with a sample of current day and some previous and consecutive samples, according to the windows size.

During the preliminary determination of optimum window size, several trials were performed, finding out that numbers which are multiples of seven, had better performance than other values, it could be due to the weekly cyclical variations related to routine and time of collection, referred previously on Figure 2.

From a collection of more than 30 initial features, preliminary testing was conducted, in which six relevant features were selected. It was determined that features like the genetic line of birds, stochastic variations in egg production, daily and cumulative mortality, weekly slope of the curve, and many others, don't provide a significant improvement to the model, and were discarded (Mollazade et al., 2012).

The feature selection for the input patterns of the SVM was defined as follows:

A. The production percentage over a day (number of eggs produced over a day/number of existing birds) minus the percentage of historical production for a similar day.

B. The production percentage over the day at the end of the sliding window, minus the production percentage over the day at the beginning of the sliding window.

C. The production over the day minus the production from seven days earlier.

D. The coefficient of variation (standard deviation / mean * 100) of the second half of the sliding window.

E. The standard deviation of the first half of the sliding window minus the standard deviation of the second half of the sliding window.

F. Age of birds in weeks.

Relevant features, as determined by the authors, were selected from the sliding window, each input pattern is having a corresponding pattern in the output set, which were zero or one, depending on whether the label of the day at the end of the forecasting interval was positive or negative regarding the presence of a problem in the curve. This procedure is performed for each day during the study period, always extracting the same fixed features.

To assess the forecasting interval, expected outputs for each sliding window has been taken from corresponding pattern in output set (zero-day forecasting interval), and a time shift (Lindsay and Cox, 2005) of one to five days later, that way SVM leaning is based next days expected outputs, and thus SVM trained could be able to detect problems prior to experts' criteria.

A k-fold cross-validation technique was used in order to ensure that the results were independent of the partition between the training and test data, also cross-validation prevents an overfitting problem (Hsu et al., 2003), thereby the subsets of each fold were a representative sample containing flocks which presented problems and flocks which did not, in a random and stratified manner.

During the k-fold cross-validation process, the data are divided into k subsets; one is used as a test subset and the others (k-1) as training subsets (Mucherino et al., 2009). The cross-validation process is repeated for k folds, with each of the possible subsets, and finally an arithmetic mean of the results for each fold is performed to obtain a single result, which is passed on to the SVM.

Thus, 100 repetitions of k-fold 5 cross-validation were performed. For this study, 12,500 support vector machines were evaluated, 500 for each factor of variation.

*2.4 Performance analysis*

The first performance requirement for a classification model is that the model generalizes well, in the sense that it provides the correct predictions for new, unseen data instances (generalization). This behavior is typically measured by percentage correctly classified test instances (accuracy), other measures include sensitivity and specificity, which are generated from a confusion matrix (Martens and Baesens, 2010).

The accuracy value is usually the only performance requirement used for evaluating the performance of machine learning techniques; this accuracy value is a statistical measure used to determine whether a binary (true or false) classification test is able to correctly identify or exclude a condition (Martens and Baesens, 2010; Venkatesan et al., 2013).

Considering that there are only 188 positive labels and 9954 negative labels in the database, samples Tang et al., (2009) states it is required to evaluate other metrics such as specificity and sensitivity, to avoid misinterpretations when having rare positive labels. A common used example to support this statement is that a classifier, which predicts all samples as negative, has high accuracy, but it is useless to detect rare positive.

The aim of this study is related to detection of problems in egg production. Therefore it is very important to achieve a highly effective detection ability for positive labels, for this, Tang et al., (2009) suggests another metric, called precision or positive predictive value.

Specificity is the ability to detect the absence of problems as false; sensitivity is the ability to detect the presence of problems as true; and positive predictive value is the probability that a problem actually occurs when the test is positive (Altman and Bland, 1994; Tang et al., 2009; Hastie et al., 2009; Venkatesan et al., 2013).

Analysis of Variance (ANOVA) and Multiple Range Tests (MRT) with Tukey's Honest Significant Difference (HSD) method for a value of $p < 0.01$, were performed to select the optimal model configuration; a positive selection of those parameter that provided the best performance metrics was carried out. Metrics were calculated from the confusion matrix of the test subset, that is, data different from those used for training, reducing the possibility of overtraining and improving its ability to generalize.

## 3. Results
### 3.1 Kernel selection

With fixed values of parameter C to 0.1, windows size to seven and forecasting interval, to one, until they are assessed respectively, four kernels were exhaustively evaluated: 1) polynomial, 2) radial basis function (RBF), 3) quadratic and 4) linear. The results of the kernel evaluation is shown in a diagram of boxes on Figure 5.

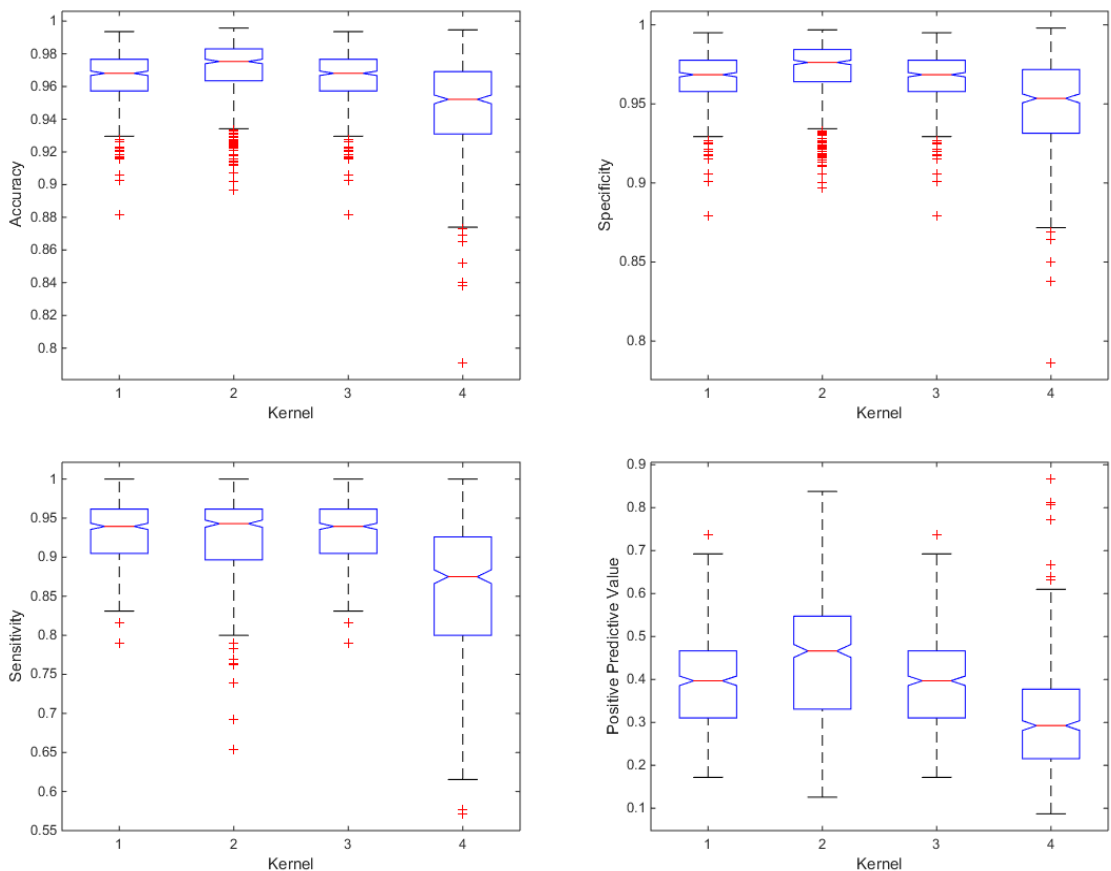

*Figure 5  Diagrams of boxes: performance metrics of the kernels evaluated.*

As shown, kernel 4 (linear) produces the worst results, whereas kernels 1, 2 and 3 (polynomial, RBF and quadratic) obtain similar results between them. An ANOVA statistical test was performed using the multiple comparison procedure, Tukey HSD, which is shown in Table 2.

Table 2 Multiple comparison (MC) of kernels for each performance metric.

| | Kernel | | | |
|---|---|---|---|---|
| | 1 polynomial | 2 radial basis function | 3 quadratic | 4 linear |
| Accuracy | 0.9654 [a] | 0.9687 [a] | 0.9654 [a] | 0.9475 [b] |
| Specificity | 0.9661 [a] | 0.9696 [a] | 0.9661 [a] | 0.9492 [b] |
| Sensitivity | 0.9289 [a] | 0.9203 [a] | 0.9289 [a] | 0.8546 [b] |
| Positive Predictive Value | 0.3932 [b] | 0.4445 [a] | 0.3932 [b] | 0.3045 [c] |

*Rows with different letters differ significantly according to Tukey's Honest Significant Difference method for a value of $p < 0.01$.*

From the above-mentioned analysis, it is selected radial basis function kernel, as it is statistically better in all four parameters evaluated, polynomial and quadratic kernels have similar performance, but have a statistically significant lower positive predictive value.

From preliminary tests performed, the sigma tuning best range was between one and six; exhaustive test through this range was performed in a gradient ascent optimization, seeking optimal performance of the model. Figure 6 shows the results of performance metrics according to values of tuned sigma.

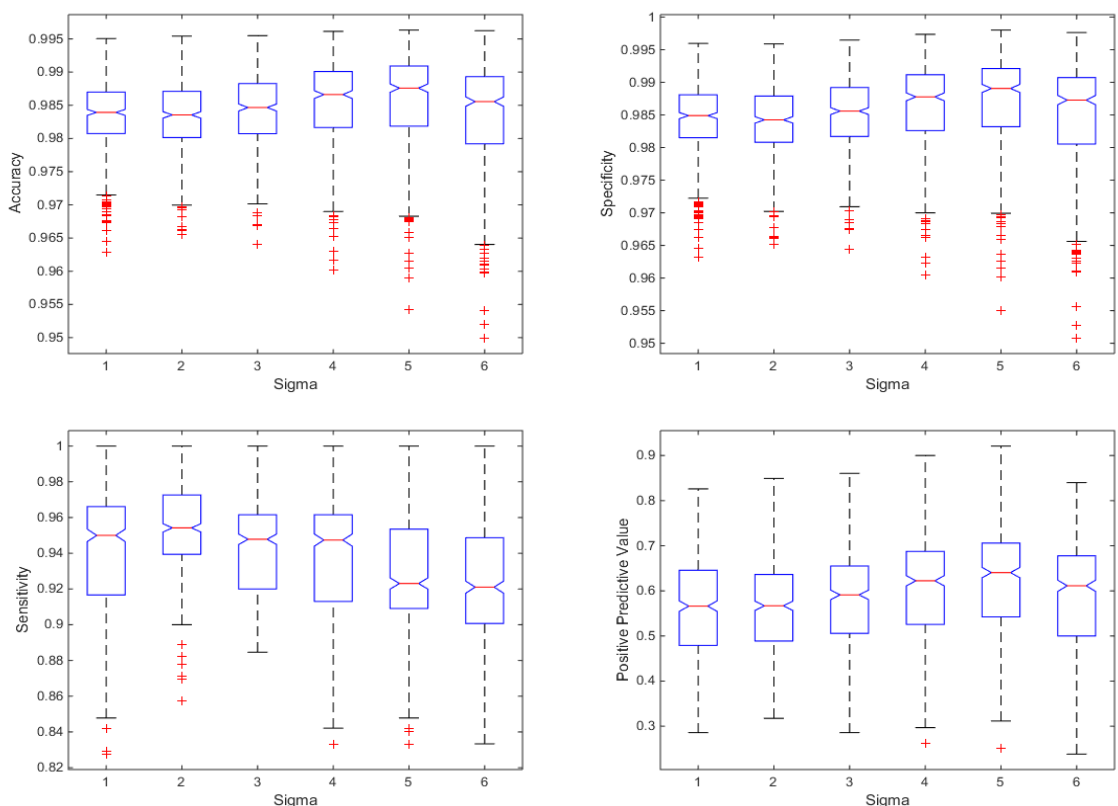

*Figure 6 Diagrams of boxes: performance metrics according to the value of sigma.*

As noted, in terms of accuracy, specificity and positive predictive value, the values tend to improve as sigma is higher (up to five), however, the sensitivity values tend to worsen as sigma is higher. In order to produce the best decision-making tool, an ANOVA statistical test was performed using the multiple comparison procedure, Tukey HSD, which is shown in Table 3.

*Table 3 MC of different sigma (σ) values for each performance metric.*

|  | σ = 1 | σ = 2 | σ = 3 | σ = 4 | σ = 5 | σ = 6 |
|---|---|---|---|---|---|---|
| Accuracy | 0.9833 [b] | 0.9833 [b] | 0.9842 [ab] | 0.9853 [a] | 0.9856 [a] | 0.9833 [b] |
| Specificity | 0.9843 [b] | 0.9840 [b] | 0.9851 [b] | 0.9864 [ab] | 0.9869 [a] | 0.9847 [b] |
| Sensitivity | 0.9406 [b] | 0.9520 [a] | 0.9419 [b] | 0.9365 [b] | 0.9259 [c] | 0.9199 [c] |
| Positive Predictive Value | 0.5657 [b] | 0.5647 [b] | 0.5836 [b] | 0.6100 [ab] | 0.6223 [a] | 0.5858 [b] |

*Rows with different letters differ significantly according to Tukey's Honest Significant Difference method for a value of p <0.01.*

A value of sigma equal to two performed the best sensitivity; however, the accuracy, specificity and positive predictive value metrics, are on group b according to Tukey's test. On the other hand, a value of sigma equal to five performed the best accuracy, specificity and positive predictive value, with the lowest sensitivity among those evaluated. A sigma value equal to five is set since it improves most of the performance metrics.

### 3.2    Parameter C

Once it was decided to set the RBF kernel, with a sigma value equal to five, an evaluation was conducted by varying the parameter C; the initial values of the window size and the forecasting interval remained constant, and only values of C below 0.25 were tested. Figure 7 shows the results from the evaluation of different values of the parameter C.

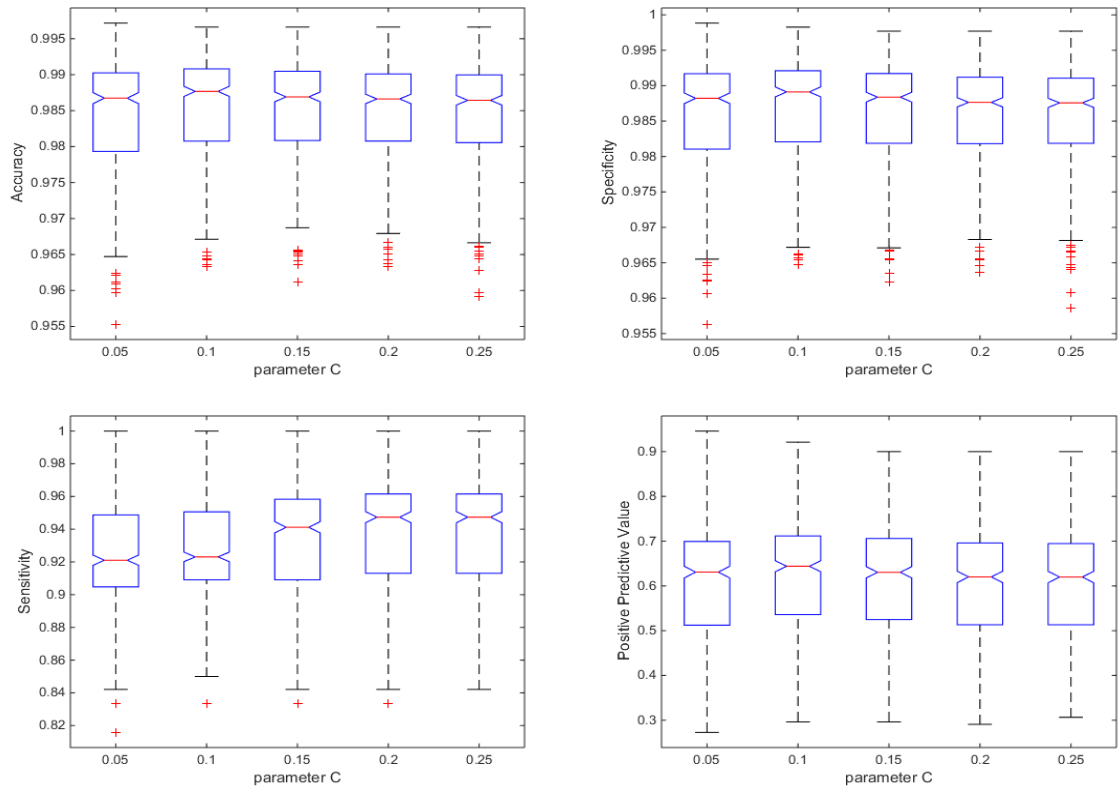

*Figure 7 Diagrams of boxes: performance metrics according to the value of the parameter C.*

As noted, for accuracy, specificity and positive predictive value, minor modifications of the parameter C do not generate significant differences, whereas, for sensitivity, the modification of this parameter does generate slight increases. In order to make the best decision, an ANOVA statistical test was performed using the multiple comparison procedure, Tukey HSD, which is shown in Table 4.

*Table 4 MC of different values of the parameter C for each performance metric.*

|  | C = 0.01 | C = 0.1 | C = 0.15 | C = 0.2 | C = 0.25 |
| --- | --- | --- | --- | --- | --- |
| Accuracy | 0.9845 a | 0.9856 a | 0.9852 a | 0.9849 a | 0.9847 a |
| Specificity | 0.9859 a | 0.9869 a | 0.9864 a | 0.9860 a | 0.9858 a |
| Sensitivity | 0.9185 b | 0.9256 b | 0.9328 a | 0.9343 a | 0.9358 a |
| Positive Predictive Value | 0.6063 a | 0.6222 a | 0.6131 a | 0.6057 a | 0.6025 a |

*Rows with different letters differ significantly according to Tukey's Honest Significant Difference method for a value of p <0.01.*

Based on the results, any value of the parameter C could be set to 0.15 or higher. Recognizing that by setting a lower value there was less possibility of overfitting, it was decided to select the parameter C value at 0.15.

## 3.3 Window size

The window size expresses the amount of data in the days before the event, which are supplied to the model in order to configure the input patterns, multiples of 7 values from 7 to 28 days were evaluated, and the results are shown below in Figure 8.

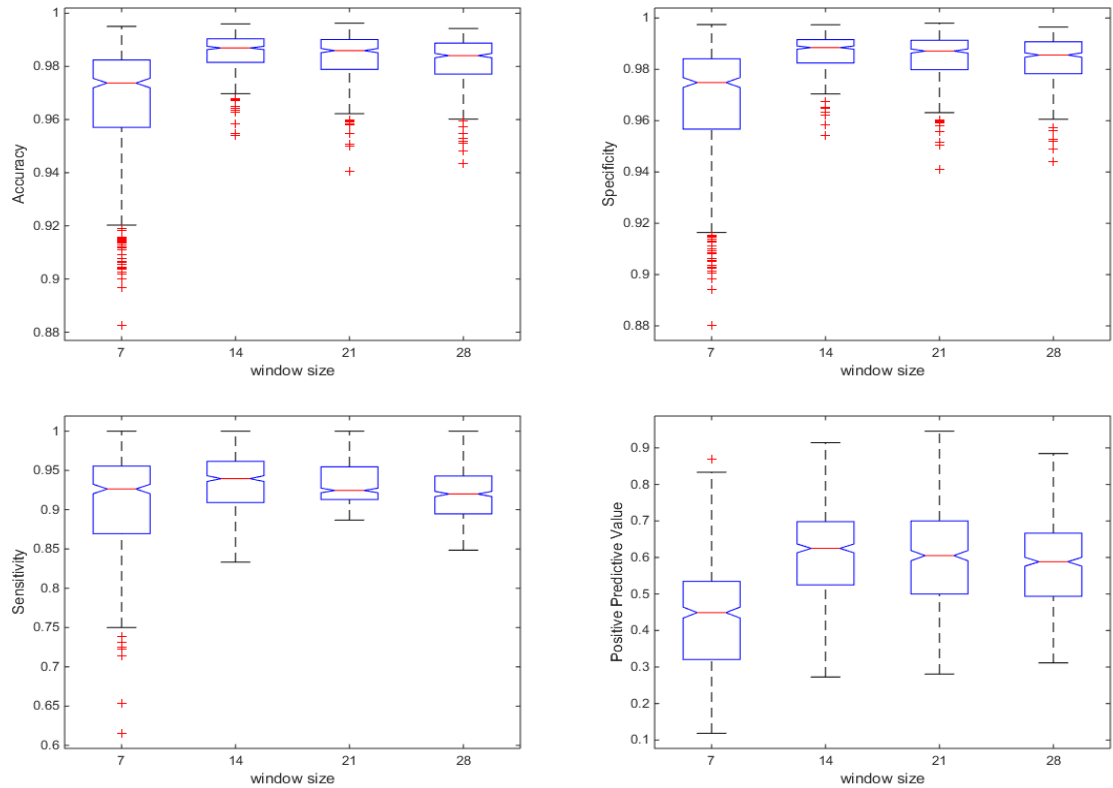

*Figure 8 Diagrams of boxes: performance metrics according to the window size.*

It is clearly noted that a window size equal to 7 generates the worst results in all the performance metrics, whereas the values among those for 14, 21 and 28 produce similar results. Therefore, an ANOVA statistical test was performed using the multiple comparison procedure, Tukey HSD, which is shown in Table 5.

*Table 5 MC of different values of window size for each performance metric.*

|  | WS = 7 | WS = 14 | WS = 21 | WS = 28 |
|---|---|---|---|---|
| Accuracy | 0.9659 c | 0.9852 a | 0.9838 ab | 0.9821 b |
| Specificity | 0.9670 c | 0.9864 a | 0.9850 ab | 0.9836 b |
| Sensitivity | 0.9020 c | 0.9318 a | 0.9320 a | 0.9168 b |
| Positive Predictive Value | 0.4300 c | 0.6122 a | 0.5972 ab | 0.5808 b |

*Rows with different letters differ significantly according to Tukey's Honest Significant Difference method for a value of p <0.01.*

The Tukey's test results unequivocally indicate that the optimal window size for this type of problem is 14 days.

## 3.4 Forecasting interval

The forecasting interval can be adjusted to suit the specific demands, a value equal to zero implies that the model works as an early warning; values higher or equal to one imply that it works as a forecasting model. Forecasting interval values were evaluated between zero and five, the results of experiments performed are shown in Figure 9.

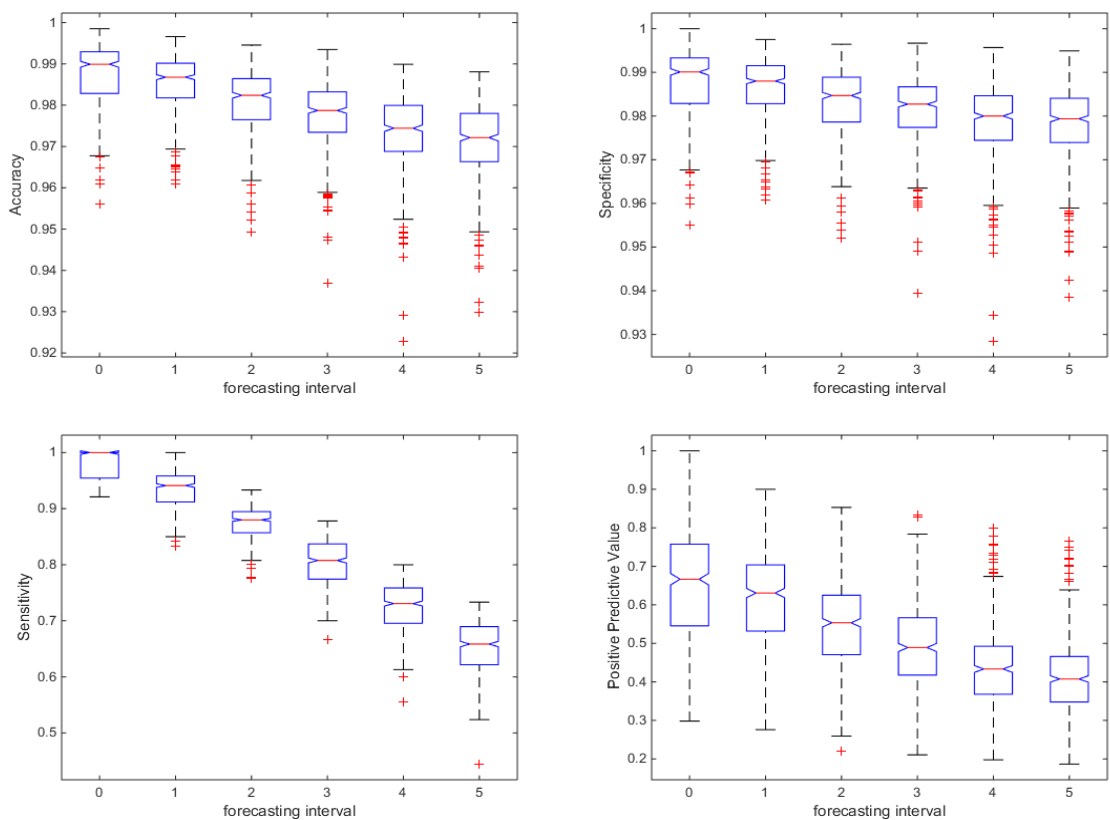

*Figure 9 Diagrams of boxes: performance metrics according to the forecasting interval.*

As expected, the shorter the forecasting interval, better performance is obtained for all performance metrics. Table 6 shows performance metrics of the assessed forecasting intervals, an ANOVA statistical test using the multiple comparison procedure, Tukey HSD was performed.

*Table 6 MC of different values of forecasting interval for each performance metrics.*

|  | FI = 0 | FI = 1 | FI = 2 | FI = 3 | FI = 4 | FI = 5 |
|---|---|---|---|---|---|---|
| Accuracy | 0.9874 a | 0.9854 b | 0.9811 c | 0.9776 d | 0.9735 e | 0.9713 f |
| Specificity | 0.9876 a | 0.9865 a | 0.9835 b | 0.9814 c | 0.9789 d | 0.9783 d |
| Sensitivity | 0.9783 a | 0.9333 b | 0.8738 c | 0.8030 d | 0.7229 e | 0.6483 f |
| Positive Predictive Value | 0.6518 a | 0.6135 b | 0.5480 c | 0.4940 d | 0.4419 e | 0.4090 f |

*Rows with different letters differ significantly according to Tukey's Honest Significant Difference method for a value of p <0.01.*

## 4. Discussion

For an early warning of problems in egg production curve, SVM classifier is proposed by authors not to classify but to detect abnormal instances, as stated by Bennett and Campbell, (2000) about novelty or abnormality detection potential applications in many problem domains. Lindsay and Cox, (2005) state that traditional machine learning techniques, like SVM, can be a viable alternative to the classical time-series analysis technique. In this study, different settings of SVM parameters were assessed using ANOVA statistical tests and Tukey Multiple Comparison tests for a value of p <0.01.

Since kernel is arguably the most important component of SVM algorithm (Suttorp and Igel, 2007; Zhao et al., 2010; Mollazade et al., 2012), exhaustive tests with four kernels were assessed in order to select the one with the best performance as proposed by Mucherino et al., (2009).

RBF, polynomial and quadratic kernels had similar performance on accuracy, specificity and sensitivity, the positive predictive value achieved by RBF kernel was better than the other kernels evaluated. The authors selected RBF kernel, which has been proved to be an excellent kernel function for several applications, agreeing with Fernández Pierna et al., (2006), Han et al., (2007), Zhao et al., (2010) and Zhiliang et al., (2015).

According to Bennett and Campbell, (2000) and Zhao et al., (2010) when RBF kernel is used, sigma parameter must be optimized, in order to obtain better performance. A common technique for this is stepping through a range of values for sigma, in a gradient ascent optimization (Suttorp and Igel, 2007). The selected range to evaluate the model was one to six.

A value of sigma equal to five performed the best accuracy, specificity and positive predictive value, 0.9856, 0.9869, 0.6223 respectively, nevertheless, performed a sensitivity value of 0.9259, the worst among those evaluated; the best sensitivity value was reached when sigma is equal to two, but in this case, the specificity value was of 0.9840.

Since the database of production of eggs, has much more negative labels than positive ones, the specificity metric has more impact on misclassifications; from this approach, a value of sigma equal to five is better. Another approach to support this decision is stated by Fernández Pierna et al., (2006) who argue that the generalization ability increases while sigma gets higher values.

Modification of the parameter C generates slight increases for sensitivity, and minor changes for the rest of metrics. Given that high values of parameter C, can cause overfitting problems (Mucherino et al., 2009), a value of 0.15 was selected since it is the lowest value with higher sensitivity performance, among the evaluated.

Window size refers to the amount of data needed by the model to perform the classification task. Besides relevant features B, D and E, depends on the amount of data provided in order to calculate a single value for each feature, which constitutes a part of a pattern.

Our results showed that a window size equal to 14 generates the best results in all the performance metrics. A windows size of 7 days, did not provide enough data, consequently patterns differ among same labels. A windows size of over 28 days, grouped excessive data, thus patterns become similar between positive and negative labels.

Forecasting interval was assessed, in a value range from zero to five, the model performed an accuracy of 0.9874, specificity of 0.9876, sensitivity of 0.9783 and a positive predictive value of 0.6518, at a forecasting interval of zero, in this case, the model works as an early warning.

As the forecasting interval increases, the performance metrics decreases, in the case of the sensitivity, the forecasting interval affects it more intensely than to other metrics. In the authors' opinion, sensitivity values above 0.8 are acceptable. Therefore, the optimal forecasting interval is considered to be from zero to three days.

At optimal forecasting interval values, the model is able to identify the problem before it became apparent to the experts' judgement. The selection of either value will depend on how accurate, sensitive and specific the model is expected to perform.

In some instances, it was found that the model was able to detect as false positives, some days prior to an event occurring. Yet those days remained overlooked by the experts as no significant reduction had been observed.

## 5. Conclusions

In this work, optimal parameter configuration of an SVM classifier model is assessed by performance metrics, results clearly indicate that it is achievable to early warn problems in the curve of commercial laying hens.

Radial basis function kernel with a sigma value equal to 5, and a parameter C value of 0.15 is the one which achieved the best performance, that is 0.9874 for accuracy, 0.9876 for specifici-ty, 0.9783 for sensitivity and 0.6518 for positive predictive value, as early warning at 0-day forecasting interval.

For this application, a window size equal to 14 generates the best results in all the performance metrics, by the modification of computed values of relevant features B, D and E, been part of input patterns.

It should be pointed out that the model has the ability to issue an alert with a sensitivity of 0.9333, 0.8738, and 0.8030, for one, two and three days respectively, before experts realized the drop of the production, the sensitivity decreases below 0.8 for greater forecasting intervals.

At farm level, an alert a day in advance, could be very helpful to decide performing a preventive diagnosis looking for clinical symptoms, or any other related issue in order to take actions for solving immediately.

## 6. Future Developments

Future work is focusing on the use of these techniques to identify features that allow for early warning of specific poultry diseases, for which a new field with confirmed diagnosis can be included in the database. Time of egg collection, daily water and food consumption, sound patterns and thermal infrared images of the birds, could be added as fields to the database in order to improve the accuracy over longer intervals of time.

The early warning model, could be embedded in hardware or production management information software, and may have a major positive impact on the poultry industry, as it allows detecting and acting in time, and could reduce economic losses related to delayed treatments.

## 7. Acknowledgements

We gratefully acknowledge to DINTA-UTMACH, RNASA-UDC and Agrolomas CL, for providing all the resources for this research; our special thanks to the two anonymous reviewers whose suggestions helped to improve and clarify this manuscript.